\documentclass[aps, reprint, superscriptaddress, prb, preprintnumbers, amsmath, amssymb, letterpaper, floatfix, showpacs]{revtex4-2}
\usepackage{amsmath, amssymb, amsfonts}
\usepackage{graphicx}
\usepackage{times}
\usepackage{dcolumn}
\usepackage{multirow}
\usepackage[colorlinks, linkcolor=blue, anchorcolor=blue, urlcolor=blue, citecolor=blue, pdfborder=001]{hyperref}

\newcommand{\NbGe}{NbGe$_2$}

\newcommand{\musr}{$\mu$SR}
\newcommand{\Tc}{$T_{\rm{c}}$}

\begin{document}


\title{\musr~study on noncentrosymmetric superconductor NbGe$_{\mathbf{2}}$}

\author{J. C. Jiao}
\author{K. W. Chen}
\affiliation{State Key Laboratory of Surface Physics, Department of Physics, Fudan University, Shanghai 200438, People's Republic of China}
\author{Adrian D. Hillier}
\affiliation{ISIS Facility, STFC Rutherford Appleton Laboratory, Harwell Science and Innovation Campus, Didcot OX11 0QX, United Kingdom}
\author{T. U. Ito}
\affiliation{Japan Atomic Energy Agency, Tokai-Mura, Ibaraki 319-1195, Japan}
\author{W. Higemoto}
\affiliation{Japan Atomic Energy Agency, Tokai-Mura, Ibaraki 319-1195, Japan}
\affiliation{Tokyo Institute of Technology, Ookayama, Meguro, 152-8550, Tokyo}
\author{Zheng Li}
\author{Baijiang Lv}
\affiliation{School of Physics, Zhejiang University, Hangzhou 310058, China}
\author{Zhu-An Xu}
\affiliation{School of Physics, Zhejiang University, Hangzhou 310058, China}
\affiliation{State Key Laboratory of Silicon and Advanced Semiconductor Materials, Zhejiang University, Hangzhou 310027,China}
\author{Lei Shu}
\altaffiliation[Corresponding Author: ]{leishu@fudan.edu.cn}
\affiliation{State Key Laboratory of Surface Physics, Department of Physics, Fudan University, Shanghai 200438, People's Republic of China}
\affiliation{Shanghai Research Center for Quantum Sciences, Shanghai 201315, People`s Republic of China}
\date{\today}

\begin{abstract}
We report a muon spin relaxation ($\mu$SR) study on polycrystalline noncentrosymmetric superconductor NbGe$_2$~with the superconducting transition temperature $T_c=2.0\sim2.1$~K. Zero-field $\mu$SR~experiment indicates the absence of spontaneous magnetic field in the superconducting state, showing the preservation of time-reversal symmetry in the superconducting state. Transverse-field $\mu$SR~experiment is performed to map the phase diagram of NbGe$_2$, from which clear evidence of both type-I and type-II superconductivity is obtained. More importantly, we clearly delineate the region in the phase diagram where type-I and type-II superconductivity coexist.  
\end{abstract}


\maketitle

\section{Introduction}

For unconventional superconductors, symmetries in addition to the U(1) gauge symmetry are broken in the superconducting state, leading to exotic behaviors and the possibility of more than one superconducting phase~\cite{Sigrist1991, Tsuei2000, Yip1993, Luke1998}. Therefore, the study of superconductors with reduced structural symmetry has been one of the central topics in superconductivity research. Among them, superconductors with noncentrosymmetric (NCS) structures are of particular interest~\cite{Smidman2017, Shang2022, Naoto2009}. In NCS superconductors, the lack of inversion symmetry leads to the appearance of an electronic antisymmetric spin-orbit coupling (ASOC)~\cite{Bauer2012}. A mixed singlet-triplet nature in the superconducting order parameter is induced with a sufficiently large ASOC, resulting in various novel features, such as  unconventional gap symmetry with nodes~\cite{Hirata2007, Samokhin2008}, topological superconductivity~\cite{Hyunsoo2018, Zhu2022}, and potential time-reversal symmetry breaking (TRSB)~\cite{Hillier2009, Barker2015, Singh2017}.

Like many Nb-rich superconductors~\cite{Matthias1954, Matthias1965, Morris1972}, the NCS compound \NbGe~with the superconducting transition temperature \Tc~$\simeq$~2.1 K was first synthesized in the 1970s~\cite{Remeika1978}, and its topological band structure has attracted renewed attention in recent years~\cite{Chang2018, Sato2023}. Specific heat and DC magnetization measurements of \NbGe~show that there is a type-I to type-II superconductivity crossover at low temperatures~\cite{Lv2020, Emmanouilidou2020}. In addition, the superconducting gap  in \NbGe~at $T=0.3$~K measured by he mechanical point-contact spectroscopy (MPCS) technique remains constant up to $H_{c1}\sim150$~Oe, and gradually decreases until $H_{c2}\sim 350$~Oe, indicating \NbGe~is going through a transition from a type-I to type-II superconductor at low temperatures~\cite{Zhang2021}. However, the mechanism of the type-I and type-II crossover in \NbGe~is not clear yet.  

In general, superconductors are classified as type-I and type-II according to the Ginzburg and Landau (GL) paradigm. The GL parameter is defined by the ratio of the penetration depth and coherence length $\kappa=\frac{\lambda}{\xi}$, and $\kappa=\frac{1}{\sqrt{2}}$ is the boundary of the type-I and type-II classification~\cite{Tinkham2004}. Type-I superconductor with $\kappa<\frac{1}{\sqrt{2}}$ exhibits Meissner, intermediate, and normal states in the $H$-$T$ phase diagram. Any real type-I superconductor sample with a nonzero demagnetizing factor $\eta$ enters into the intermediate state when the external field is in the range $(1-\eta)H_{c}<H_{ext}<H_{c}$. For type-II superconductors with $\kappa>\frac{1}{\sqrt{2}}$, quantized magnetic flux penetrates the sample forming Abrikosov vortices in external fields $H_{c1}<H_{ext}<H_{c2}$. A stable vortex lattice can be formed due to the repulsive interactions between the flux vortices, and such a state is known as a vortex state~\cite{Marlyse1996}. A superconductor is generally either type-I or type-II, but in a few superconductors, the coexistence of type-I and type-II superconducting signals is observed~\cite{Kortus2001, Lortz2005}, showing a unique intermediate-mixed state (IMS) in the phase diagram~\cite{Jacobs1973, Christen1980}, which is named as type-I/type-II superconductivity. 

Type-I/type-II superconductors discovered so far have multi-band characteristics. For example, MgB$_2$~\cite{Kortus2001} and ZrB$_{12}$~\cite{Lortz2005} have multiple coherence lengths and meet the conditions of $\lambda/\xi_1>1/\sqrt{2}$ and $\lambda/\xi_2<1/\sqrt{2}$, leading to the coexistence of type-I and type-II superconducting behavior~\cite{Babaev2005, Babaev2011, Babaev2012, Babaev2019}. However, there is a lack of experimental evidence that \NbGe~exhibits multi-band features. Therefore, the multi-band theory for the type-I/type-II superconductivity may not apply to \NbGe. In single-band models~\cite{Jacobs1973, Christen1980}, microscopic corrections beyond GL theory can also lead to the appearance of IMS when $\kappa\simeq1/\sqrt{2}$. However, $\kappa$ of \NbGe~estimated by specific heat and electrical resistance measurements is much smaller than $1/\sqrt{2}$~\cite{Lv2020}. Therefore, more experimental measurements  are required to understand the nature of type-I/type-II superconductivity in \NbGe.

The muon spin relaxation (\musr) technique uses muon as a local probe to detect magnetic signals in samples. Zero-field (ZF)-\musr~ experiment can detect extremely small magnetic fields, and has a great advantage in exploring the potential TRSB superconductivity in NCS superconductors. Transverse-field (TF)-\musr~experiment measures the internal magnetic field distribution, and it has been widely used to map the phase diagram, and study the microscopic properties of both type-I and type-II superconductors~\cite{Sonier2000, Beare2019, Leng2019, Karl2019, Biswas2020}. In addition, the relaxation rate measured in TF-\musr~experiment is associated with the magnetic penetration depth $\lambda$ and the GL parameter $\kappa$~\cite{Brandt1988, Sonier2000, Brandt2003}.  We report the results of ZF-\musr \ measurements performed on polycrystalline \NbGe, suggesting that the time-reversal symmetry is preserved in the superconducting state of \NbGe. Different superconducting states of \NbGe~are revealed by the TF-\musr\ measurements. A detailed $H$-$T$ diagram of \NbGe~is obtained, indicating the coexistence of type-I and type-II superconductivity in \NbGe.  In addition, the temperature dependence of the GL parameter $\kappa$ is estimated. Thus, we phenomenologically explain the appearance of type-I/type-II superconductivity in \NbGe.

\section{Experiments}

Polycrystalline samples of \NbGe~were prepared by the solid-state reaction at Zhejiang University, and the characterization data can be found in Ref. \onlinecite{Lv2020}. The definition of the critical fields $H_c$, $H_{c2}$, and $H_{c3}$ in \NbGe~follows the description in Ref.~\onlinecite{Lv2020} as well.

\musr~experiments were carried out using the MUSR spectrometer at the ISIS Neutron and Muon Facility, Rutherford Appleton Laboratory, Didcot, UK and the D1 spectrometer at J-PARC, Tokai, Japan. ZF-\musr~experiments were performed above and below \Tc~down to $T$~=~0.02~K to study whether there is a spontaneous small magnetic field in the superconducting state due to the TRS breaking. 

\begin{figure}[ht]
	\begin{center}
		\includegraphics[clip=,width=6cm]{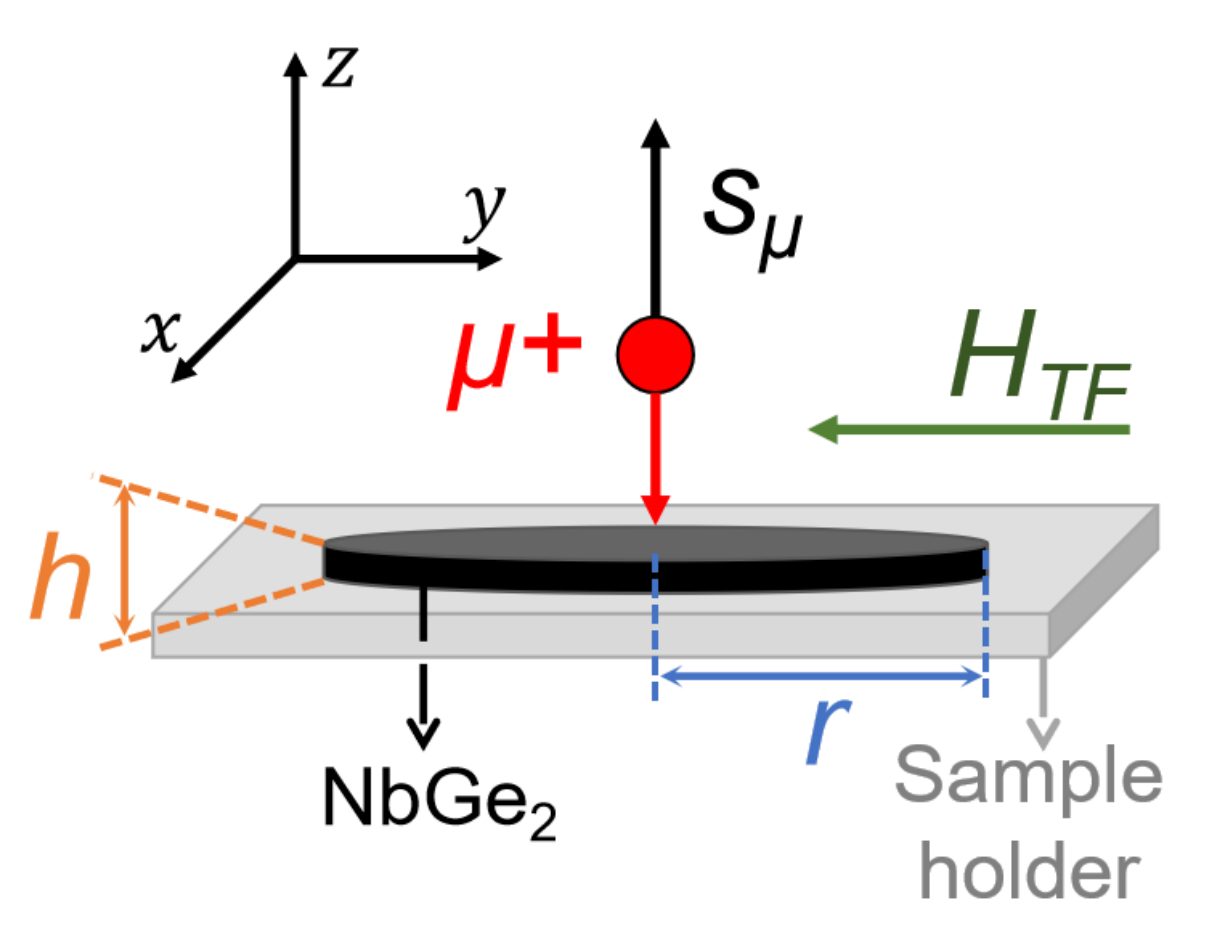}
		\caption{The schematic diagram of the TF-\musr~setup.}
		\label{fig: Exp}
	\end{center}
\end{figure}

In TF-\musr, measurements with a range of temperatures and applied magnetic fields were undertaken to map the phase diagram of \NbGe. For a type-I/type-II superconductor, the field distribution inside the sample is complicate due to the possibles of Meissner, intermediate, IMS, vortex, and normal states. In the TF-\musr~experiment, the external field $H_{ext}$ is applied perpendicularly to the initial muon spin polarization. After each muon is implanted into the sample, the muon spin precesses about the local magnetic field $B_{loc}$ at the muon stopping site with the Larmor frequency $\omega = \gamma_{\mu}B_{loc}$. By analyzing the position of the field peak and number of the field line shapes from the Fourier transformations (FFTs), the state of the sample can be better determined. In addition, if a series of FFTs curves of a $T$-scan under a fixed external field are simultaneously examined, the phase transformation process of the sample can be directly measured. Figure~\ref{fig: Exp} shows the schematic diagram of our TF-\musr~setup. The \NbGe~polycrystalline sample was mounted on a sample holder located in the $xy$ plane, and it was pressed and sintered into a shape of cylinder with the radius $r=22(1)$~mm and the height $h=1.2(2)$~mm. Muons were implanted into the sample along the $z$-axis, and $H_{TF}$ was applied parallel to the $xy$ plane.

The \musr~data were analyzed by using the MUSRFIT~\cite{Suter2012} and the MANTID~\cite{Arnold2014} software package. 

\section{Results}

\subsection{ZF-\musr}

\begin{figure}[ht]
	\begin{center}
		\includegraphics[clip=,width=8.5cm]{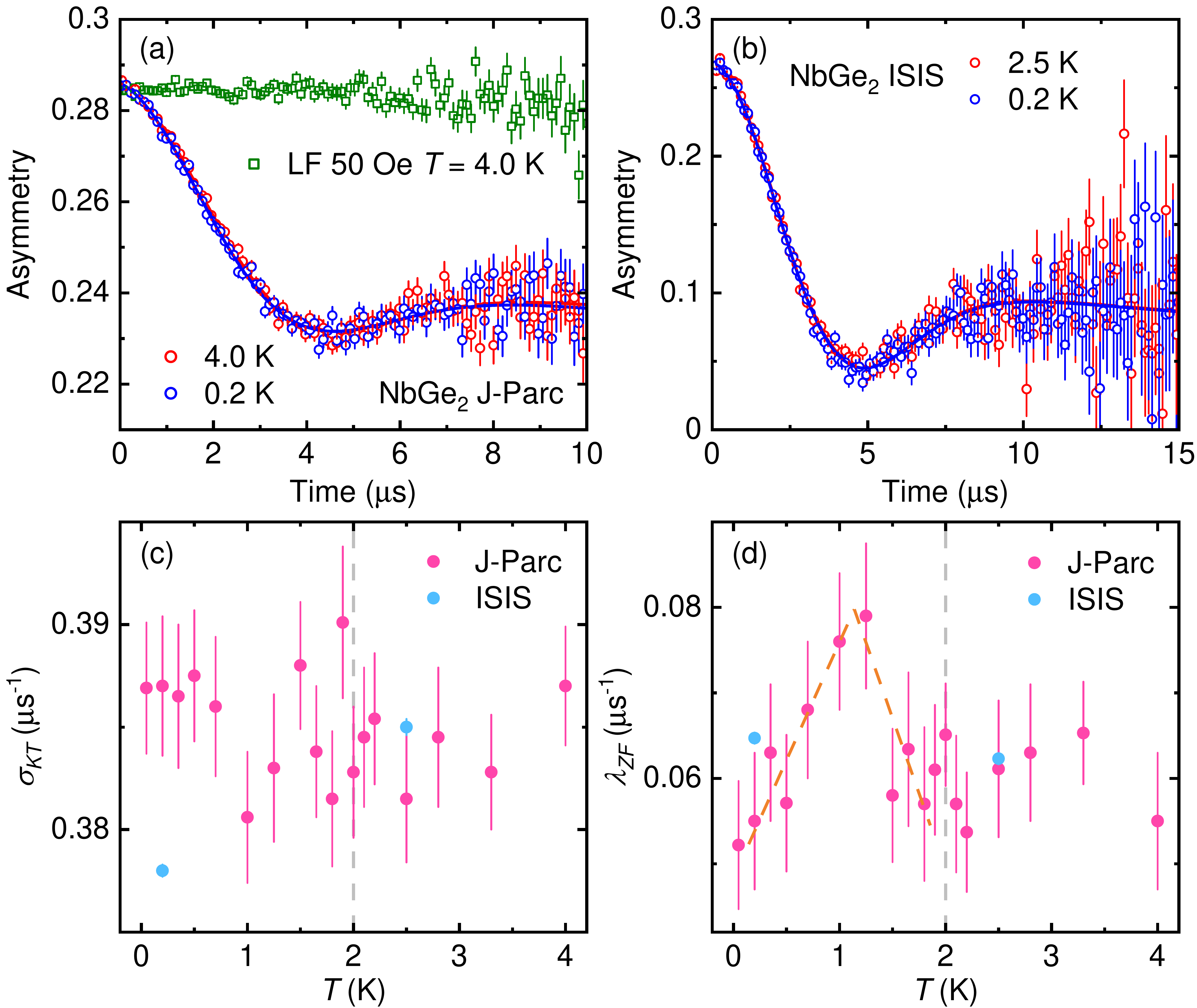}
		\caption{(a) ZF and LF \musr~asymmetry spectra collected at J-Parc. (b) ZF \musr~ asymmetry spectra collected at ISIS. In (a) and (b): Blue circles: superconducting state. Red circles: normal state. Solid curves: fits to the data with Eq.~(\ref{eq:ZF}). (c), (d) Temperature dependences of the relaxation rates $\sigma_{KT}$ and  $\lambda_{ZF}$, respectively. Grey dashed line marks \Tc. Orange dashed line is guide for the eyes. }
		\label{fig: ZF}
	\end{center}
\end{figure}

ZF-\musr~spectra at temperatures above and below \Tc~measured at J-Parc and ISIS are shown in Fig.~\ref{fig: ZF}(a) and Fig.~\ref{fig: ZF}(b), respectively. The \musr~asymmetry spectrum consists of two signal parts, which are from muons that stop in the sample and the silver sample holder, respectively. The spectra can be well fitted by the function:
\begin{equation}
	\label{eq:ZF}
	Asy(t) = A_{0}[fe^{-\lambda_{ZF}t}G_{ZF}^{KT}(\sigma_{KT}, t) + (1 - f)e^{-\lambda_{bg}t}],
\end{equation}
where $A_{0}$\ and $f$ represent the initial asymmetry and the fraction of muons stopping in the sample, respectively. The Kubo-Toyabe (KT) term~\cite{Kubo1979}
\begin{equation}
	\label{eq:KT}
	G_{ZF}^{KT}(\sigma_{KT}, t) = \frac{1}{3}+\frac{2}{3}(1-\sigma_{KT}^{2}t^{2})\exp(-\frac{1}{2}\sigma_{KT}^{2}t^{2})
\end{equation}
describes a Gaussian distribution of randomly oriented static local fields with the distribution widths $\delta B_{\rm{G}}= \sigma_{ZF}/\gamma_{\mu}$, where $\gamma_{\mu} = 2\pi \times135.53\  \rm{MHz/T}$ is the muon gyromagnetic ratio and $\sigma_{KT}$\ is the relaxation rate. 

The temperature dependence of the relaxation rate $\sigma_{KT}$ is shown in Fig.~\ref{fig: ZF}(c). No significant change crossing \Tc~is observed down to $T=20$~mK in our study. Such results indicate that there is no spontaneous magnetic field appearing in the superconducting phase. 

As shown in Fig.~\ref{fig: ZF}(a), there is little relaxation with a weak applied magnetic field $H_{LF}$ = 5 mT, indicating that the muon spin relaxation in \NbGe~is mostly due to the static or quasi-static field distributions. The exponential damping with rate $\lambda_{ZF}$ in Eq.~(\ref{eq:ZF}) at low temperatures generally stems from the rapidly fluctuating electronic spins. Fig.~\ref{fig: ZF}(d) shows the temperature dependence of the relaxation rate $\lambda_{ZF}$ in \NbGe. $\lambda_{ZF}$ shows no obvious temperature dependence in the normal state.  A peak of $\lambda_{ZF}$($T$) shows up around $T_c$, and then $\lambda_{ZF}$ decreases with decreasing temperature.  $\lambda_{ZF}$($T$) is similar to the $^{73}$Ge nuclear spin-lattice relaxation rate 1/$^{73}T_1(T)$, measured using zero-field nuclear quadrupole resonance (NQR) in PrPt$_4$Ge$_{12}$~\cite{Haller2010}, and to the temperature dependence of muon spin relaxation rate in Pr-rich samples in Pr$_{1-x}$Ce$_x$Pt$_4$Ge$_{12}$~\cite{Zhang2015} and Pr$_{1-x}$La$_x$Pt$_4$Ge$_{12}$~\cite{Zhang2019}, which resembles the Hebel-Slichter “coherence” peak expected in a superconductor with an isotropic gap~\cite{Hebel1959}.

\subsection{TF-\musr}

\begin{figure}
	\begin{center}
		\includegraphics[clip=,width=6cm]{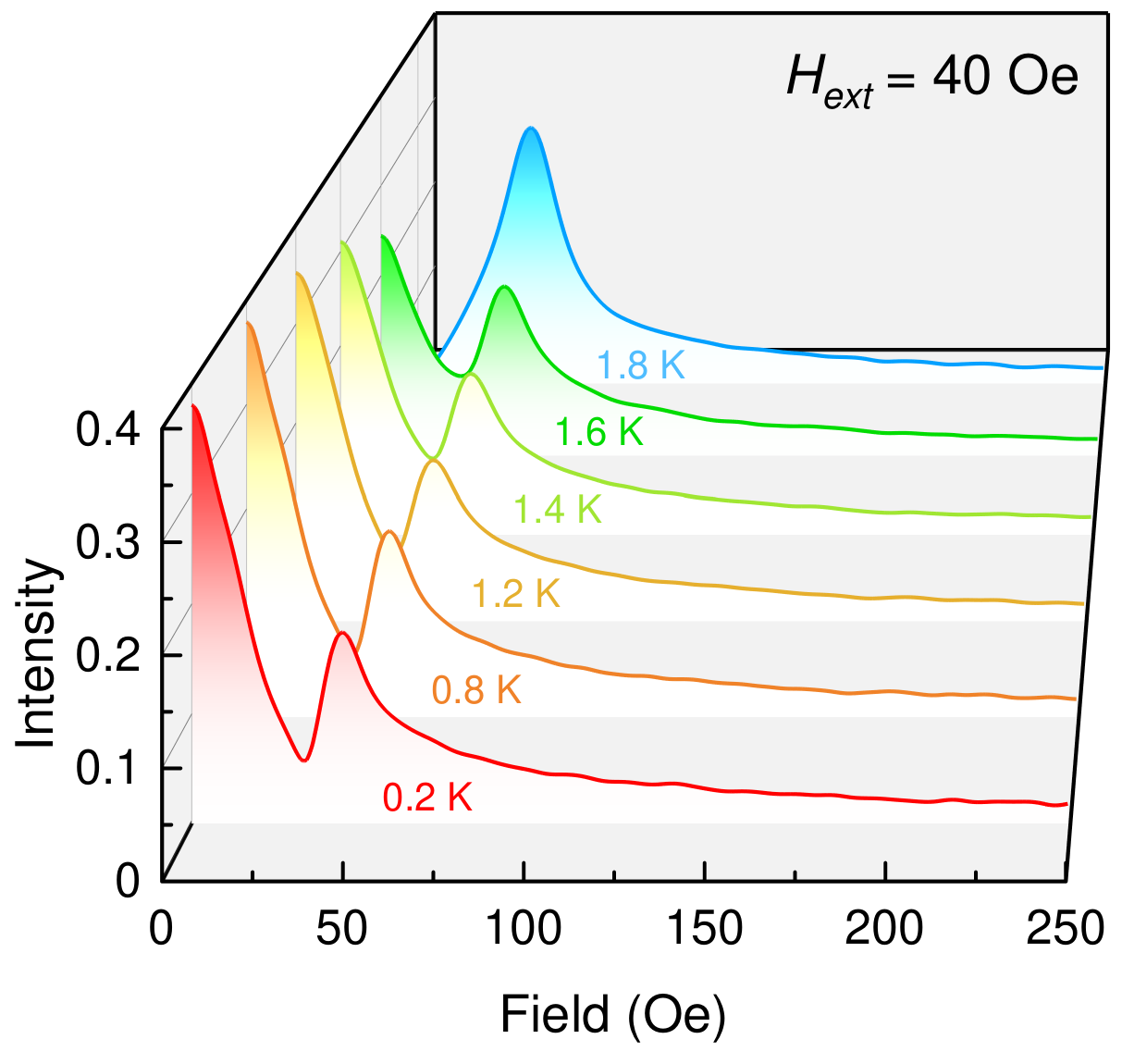}
		\caption{FFTs curves from $T = 0.2$~K to 1.8 K at the external field $H_{ext}=40$~Oe.}
		\label{fig: TF1}
	\end{center}
\end{figure}

The FFTs curves of a series of TF spectra measured at the lowest field $H_{ext}=40$~Oe are shown in Fig.~\ref{fig: TF1}. At $T=0.2$~K, $H_{ext}=40$~Oe is well below the critical field. The FFTs result shows a strong component at zero magnetic field besides the background signal. The absence of any additional magnetic signals indicates that the magnetic field is completely expelled from the sample. As the temperature gradually increases to $T=1.8$~K, no other field signals appears, indicating that the \NbGe~stays in the Meissner state before entering the normal state. The absence of the vortex-state signal is consistent with the type-I superconductivity. Thus \NbGe~shows possible type-I superconductivity with the applied field $H_{ext}=40$~Oe.

\begin{figure}
	\begin{center}
		\includegraphics[clip=,width=8.5cm]{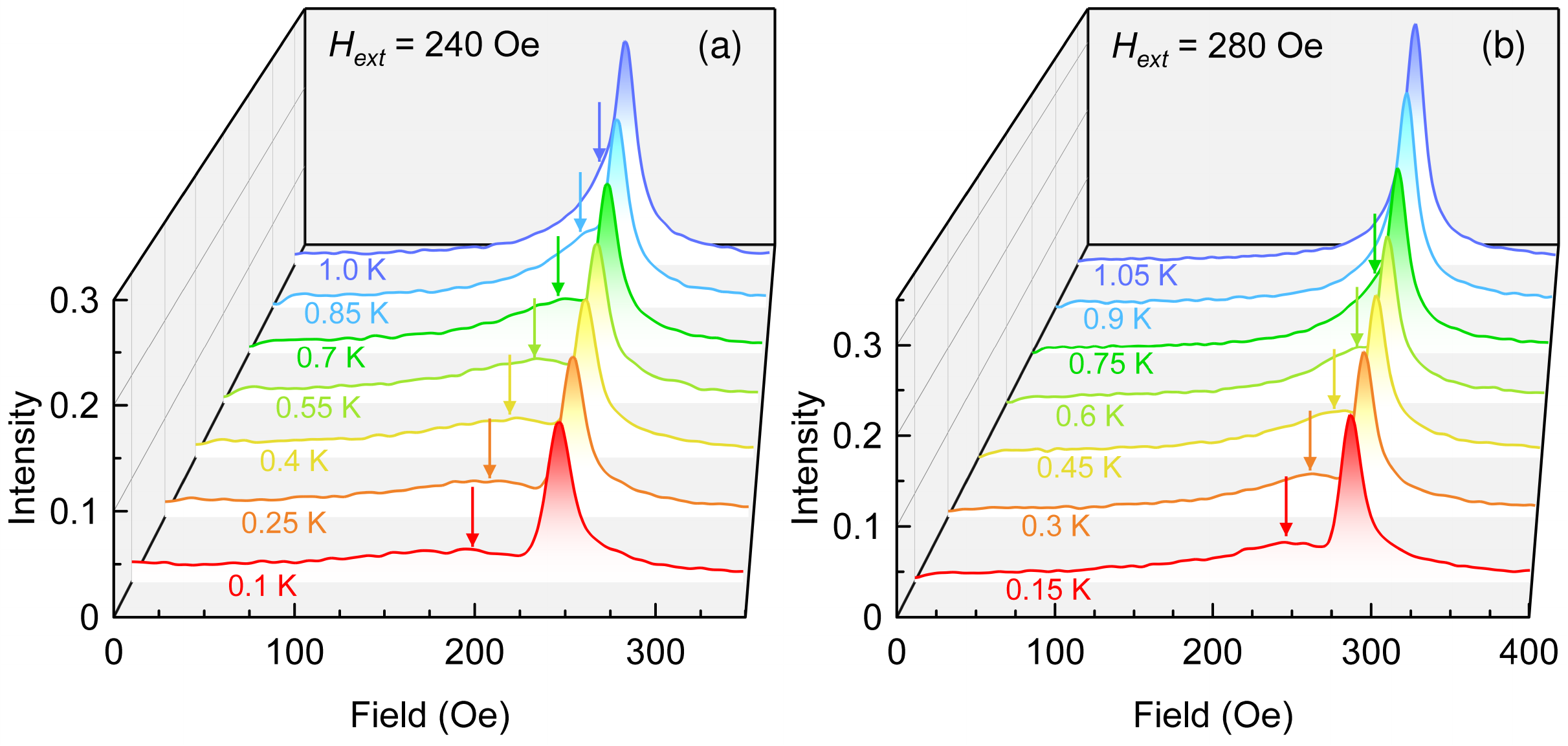}
		\caption{FFTs curves of a series of TF spectra measured at $H_{ext}=$ (a) 240 Oe and (b) 280 Oe. Arrows mark the peak of the internal field disrtibution below $H_{ext}$. The absence of zero-field peak in all these curves indicates that the sample only exhibit the type-II superconductivity.}
		\label{fig: TF2}
	\end{center}
\end{figure}

Figure~\ref{fig: TF2} shows the FFTs curves measured at larger magnetic fields ($H_{ext}>0.67H_c(0)$). At $H_{ext}=240$~Oe and 280~Oe, Gaussian distributions of fields centered below $H_{ext}$ are observed in each FFTs curve shown in Fig.~\ref{fig: TF2}. The vortex state has a distinctive feature of Type-II superconductivity and is characterized by the quantized magnetic flux lattice. The internal field distribution will form a peak known as the saddle point, which denotes the most likely field value. The arrows in Fig.~\ref{fig: TF2}~mark the position of the saddle point in each FFTs curve. As the temperature increases, the peak of the internal field gradually approaches $H_{ext}$.  Therefore, \NbGe~is found to be in the vortex state. The absence of FFTs peak at zero field suggests that the full volume of the sample is in the vortex state, exhibiting pure type-II superconductivity for $H_{ext}=240$~Oe and 280~Oe.

\begin{figure}
	\begin{center}
		\includegraphics[clip=,width=8.5cm]{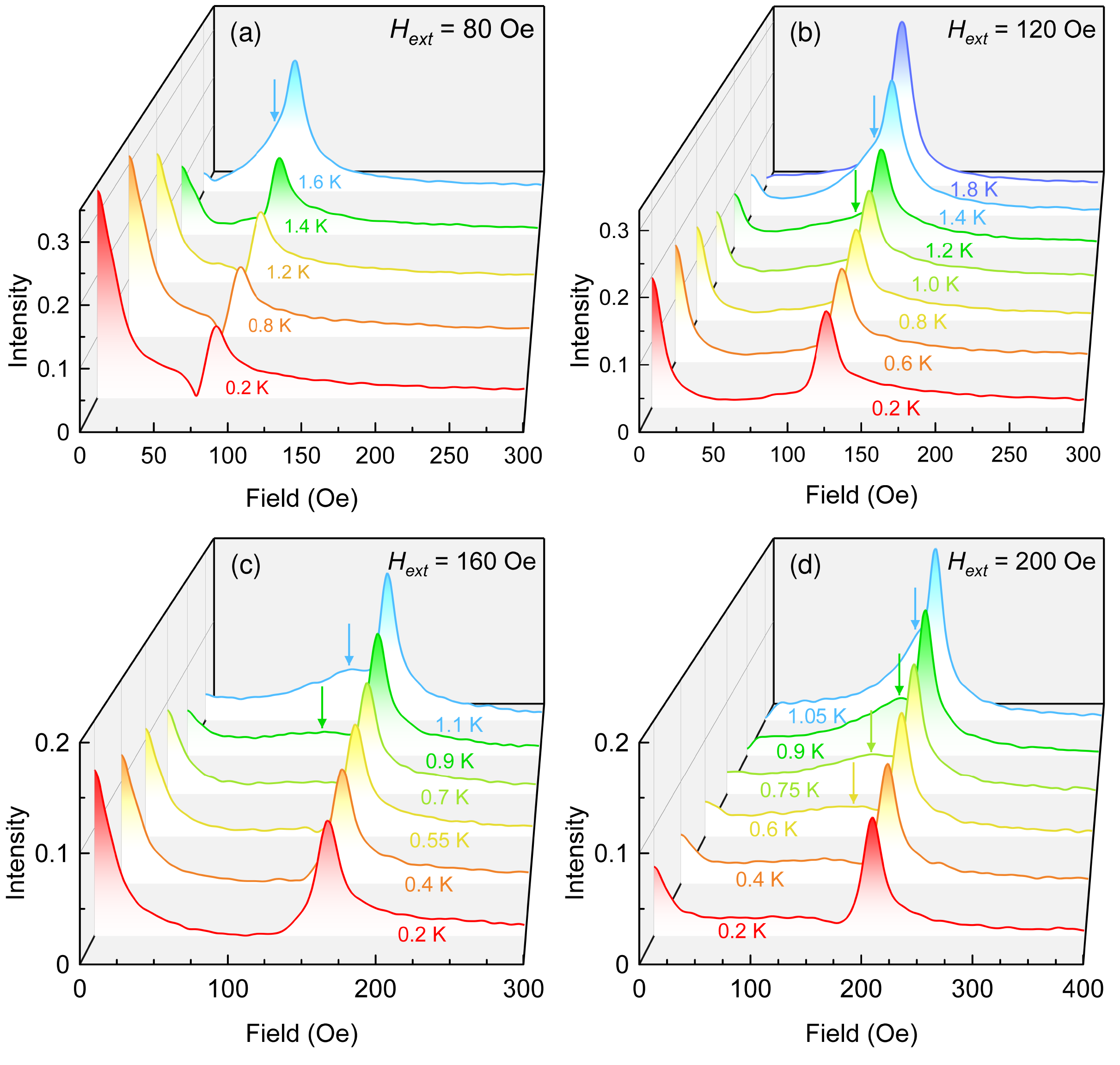}
		\caption{FFTs curves of a series of TF spectra measured at $H_{ext}=$ (a) 80 Oe, (b) 120 Oe, (c) 160 Oe, and (d) 200 Oe. Arrows mark the peak of the internal field disrtibution below $H_{ext}$.}
		\label{fig: TF3}
	\end{center}
\end{figure}

When applying $H_{ext}$ in the range from 80 Oe to 200 Oe, the field distributions inside the sample are more complicated. Fig.~\ref{fig: TF3} shows the FFTs curves measured at magnetic fields $H_{ext}=$ 80 Oe, 120 Oe, 160 Oe, and 200 Oe. For several FFTs curves shown in Fig.~\ref{fig: TF3}, such as $H_{ext}=120$~Oe $T=1.4$~K and $H_{ext}=160$~Oe $T=0.9$~K, we observe the coexistence of Meissner state and vortex state signals.  Such a state arises when vortices have weak attractive interaction. Due to the direction of the applied magnetic field parallel to the surface of the disk (Fig.~\ref{fig: Exp}), the demagnetization factor is small. Therefore, our TF-$\mu$SR data cannot intuitively reflect the IMS state, in which a third peak would appear at the critical field $H_c$ besides the Meissner and vortex-state peaks~\cite{Singh2019, Biswas2020, Jiao2024}. The coexistence of the Meissner and the vortex-state signal is consistent with a type-I/type-II superconductor~\cite{Biswas2020, Jiao2024}. In addition, the coexistence region of the Meissner and the vortex state observed in the \musr~experiment is consistent with the region where type-I/type-II superconductivity exists in \NbGe~in previous reports~\cite{Lv2020,Zhang2021,Emmanouilidou2020}. Therefore, it can be considered that the Meissner-vortex state observed from \musr~data in \NbGe~is a sign of type-I/type-II superconductivity.

In addition, for $H_{ext}=$ 80 Oe, 120 Oe, and 160 Oe (shown in Fig.~\ref{fig: TF3}(a), (b), and (c)), before entering the normal state, signals of the Meissner state gradually weaken but do not disappear. On the contrary, for $H_{ext}=$ 200 Oe (shown in Fig.~\ref{fig: TF3}(d)), the signal of the Meissner state completely disappears at $T=0.9$~K. These phenomena indicate that under a given external field, as the temperature increases, there are various evolution processes in the superconducting properties of the sample. This makes the $H$-$T$ phase diagram of \NbGe~exhibit a more complicated form than that of conventional type-I or type-II superconductors.

\begin{figure}[ht]
	\begin{center}
		\includegraphics[clip=,width=8.5cm]{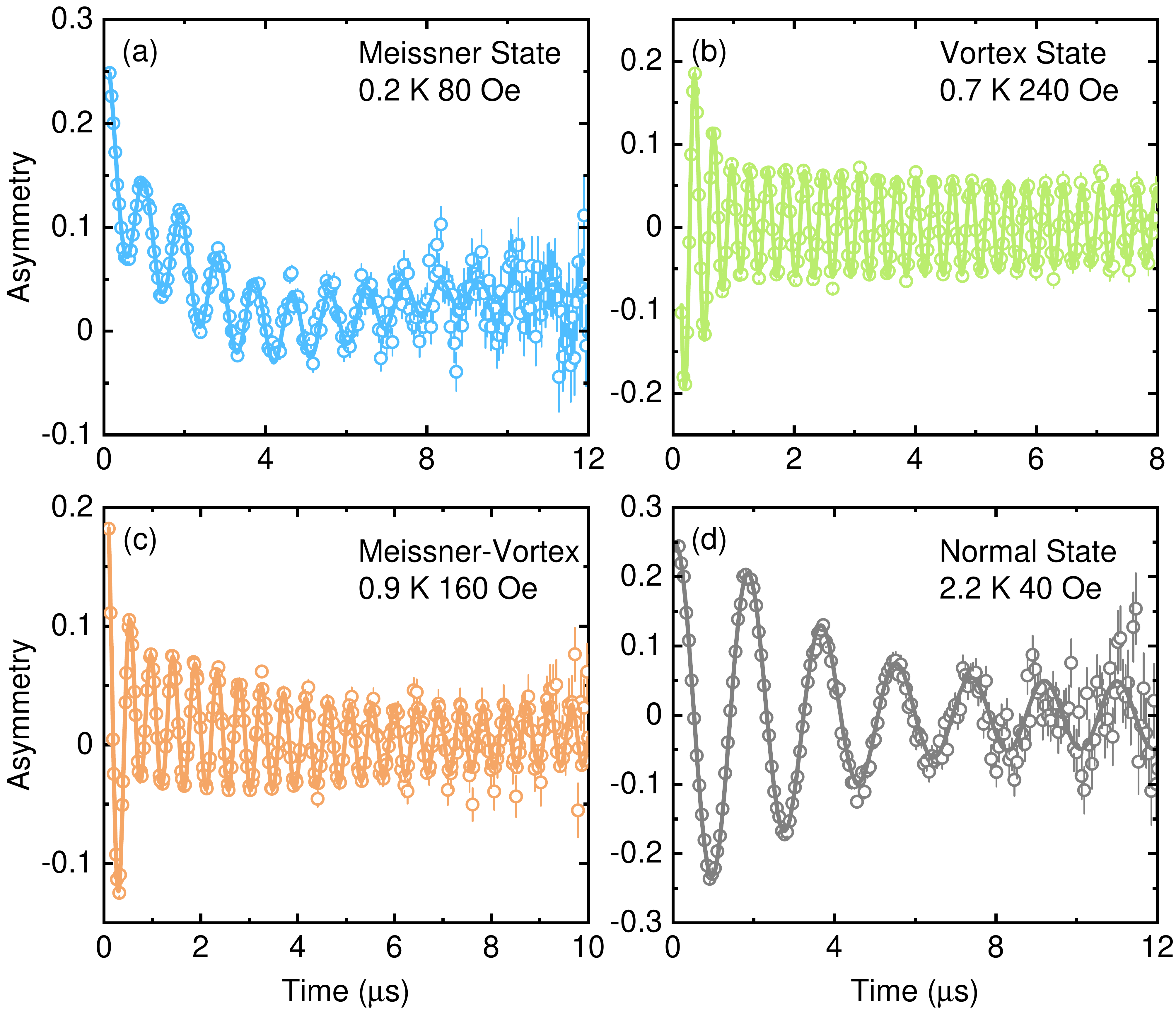}
		\caption{TF-\musr~time spectra collected at different temperatures and applied fields. The figure illustrates the typical signal observed in the (a) Meissner, (b) Vortex, (c) Meissner-vortex, and (d) Normal state. The solid curves are fits to the data using Eq.~(\ref{eq:TF}).}
		\label{fig: TF4}
	\end{center}
\end{figure}

Having the information of the field distribution, the TF-\musr~spectra can be analyzed by using the following formula:
\begin{equation}
	\begin{split}
	\label{eq:TF}
   Asy(t) = A_{bg}e^{-\lambda_{bg}t}\cos(\omega_{bg}t+\phi)\\
   + A_{Meiss}G_{ZF}^{KT}(\sigma_{KT}, t) \\
   +
   \sum_{i=vortex, normal}A_{i}\exp(-\frac{1}{2}\sigma_{i}^{2}t^{2})\cos(\omega_{i}t+\phi),
    \end{split}
\end{equation}
where the first term is due to the muons stopped into the silver sample holder, and the second term is the Kubo-Toyabe function representing the response of the static nuclear moments in the Meissner state. The relaxation rate $\sigma_{KT}$~is fixed at the same value as that in our ZF-\musr~experiment. The last term in Eq.~(\ref{eq:TF})~ includes three Gaussian distributions of non-zero magnetic fields in the vortex and normal state. It should be mentioned that all four terms in Eq.~(\ref{eq:TF})~do not exist at the same time in the fitting of one TF-\musr~spectrum. For instance, if the sample is in the vortex state, there is no Meissner contribution in its TF-\musr~spectrum. A set of TF-\musr~spectra are shown in Fig.~\ref{fig: TF4}. These representative spectra respectively show that the sample is in: (a) Meissner, (b) Vortex, (c) Meissner-vortex, and (d) Normal state. The different TF-$\mu$SR time spectra and the successful fitting using Eq.~(\ref{eq:TF}) with different line shape information confirm each state of the sample.

\section{Discussion}

\subsection{Phase Diagram}

\begin{figure}[ht]
	\begin{center}
		\includegraphics[clip=,width=8.5 cm]{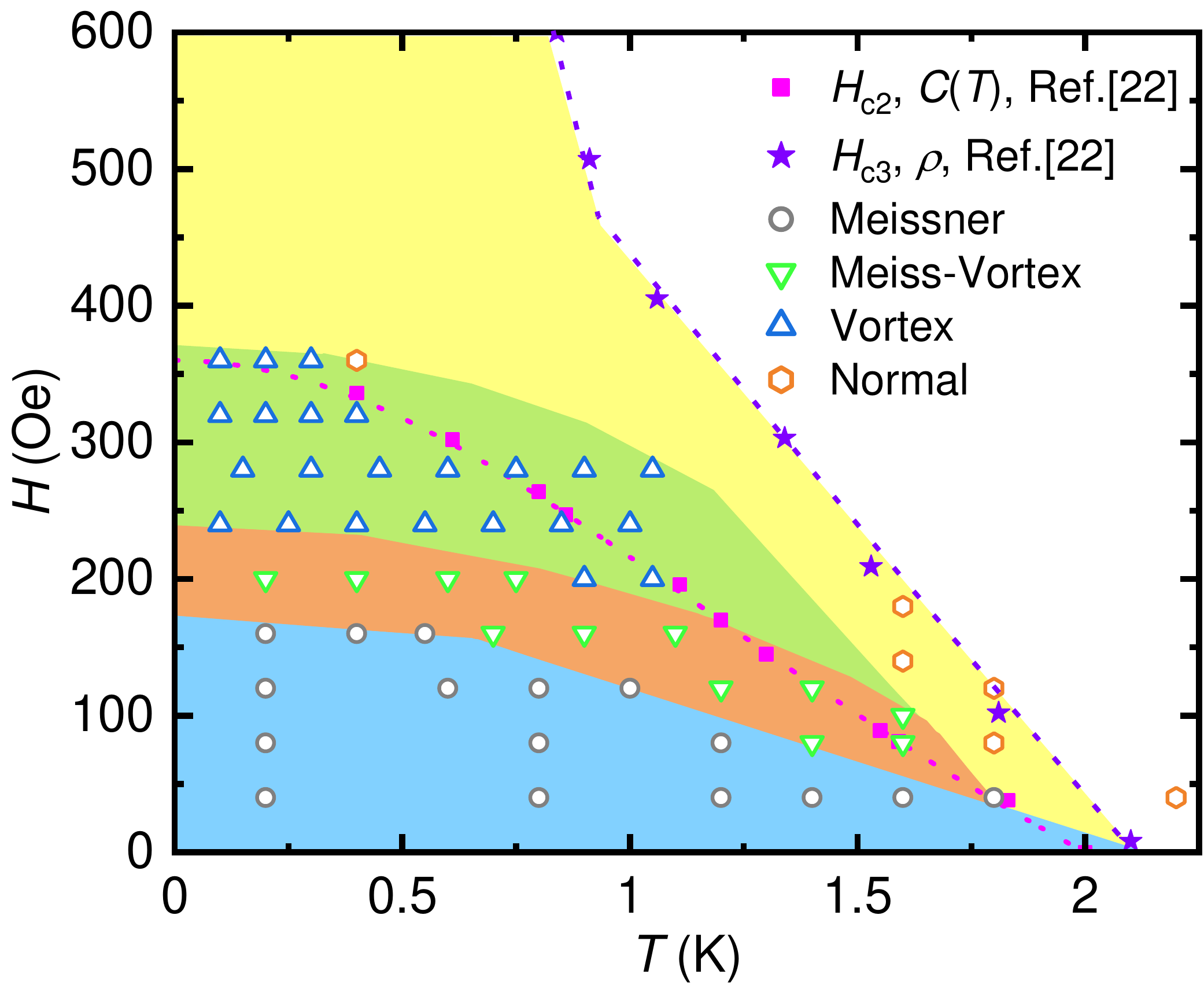}
		\caption{Different superconducting phases of \NbGe~as described in the text. Blue and green zones represent that the sample shows type-I and type-II superconductivity, respectively. The coexistence of type-I and type-II superconductivity is shown in the orange zone. Purple stars and purple dashed line represent the surface critical field $H_{c3}$ obtained from the resistivity measurement in Ref.~\cite{Lv2020}, showing as the boundary of surface superconductivity. Red square and red dashed line represent the critical field $H_{c2}$ obtained from the specific heat measurement in Ref.~\cite{Lv2020}. }
		\label{fig: PD}
	\end{center}
\end{figure}

TF-\musr~data reveals different states in \NbGe~at various fields and temperatures. These are summarized in a $H$-$T$ phase diagram in Fig. \ref{fig: PD}.

In addition to the type-I and type-II superconducting states, the unconventional state showing the coexistence of type-I and type-II superconductivity are observed. 

In \musr~studies on type-I superconductors, signals of the intermediate state, in which there is a coexistence of magnetic fields with regions of zero field and internal field $H_c$, can generally be observed under conditions of small external fields and temperatures close to \Tc~\cite{Beare2019, Leng2019}. The absence of the intermediate-state signal in this study can be understood by considering the demagnetization factor $\eta$.  In the TF-\musr~set up shown in Fig.~\ref{fig: TF1}(a), $\eta$ of the measured sample is determined by the following function~\cite{Sato1989}:
\begin{equation}
	\label{eq:DeMag}
	\eta = (\frac{h}{r\sqrt{\pi}})/[2(\frac{h}{r\sqrt{\pi}})+1]
\end{equation}
to be 3.0(5)\%. Therefore, the intermediate states can be only observed in a very narrow parameter space $0.97H_c<H_{ext}<H_c$. In our experiments, the $T$-scan points were taken at an interval of 0.1-0.2 K, it is difficult to get significant intermediate-state signals. Thus, the FFTs spectrum containing only zero field is considered to reflect type-I superconductivity. In real situations, the demagnetization factor of polycrystalline samples may be affected by the randomly arranged individual grains. And the demagnetization factor of the thin-disc shape sample may depend on the coercivity and state function of the sample and the geometrical properties of the disc. The actual demagnetization factor value of the \NbGe~sample may deviate from the value calculated by Eq.~(\ref{eq:DeMag}) by about 10\%~\cite{Brug1985, Bahl2013, Bjork2019}. However, this does not fundamentally affect the conclusions in this paper.

For \NbGe, the surface superconductivity, which is characterized by a large surface critical field $H_{c3}$, has been discussed in the previous work~\cite{Lv2020}. The surface state caused by ASOC~\cite{Manfred2014} is considered to be a certain feature of NCS superconductors~\cite{Wakui2009, Kimura2016}. The surface critical field $H_{c3}$ obtained from the resistivity measurement and the boundary of the bulk and the surface superconductivity $H_{c}$ obtained by the specific heat $C(T)$ measurement~\cite{Lv2020} are shown in Fig.~\ref{fig: PD}. 

Our \musr~data show extra bulk superconducting signals between $H_c$ and $H_{c3}$ curves in the phase diagram. It should be noted that powdered polycrystalline \NbGe~samples were used in our \musr~measurement, while single crystal samples were used in the specific heat and the resistivity measurements~\cite{Lv2020}. Polycrystalline sample typically contains more defects and impurities that provide additional centers for flux pinning, making it harder for flux vortices to move. Thus, it may exhibit larger critical magnetic fields.  On the other hand, surface states of polycrystalline powder samples may be more pronounced compared to single-crystal samples, since polycrystalline samples tend to have more grain boundaries and defects on their surfaces, leading to the formation of more surface states. Therefore, a larger critical magnetic field $H_{c2}$ was obtained in the polycrystalline sample used in our \musr~experiments.

\subsection{Type-I/type-II Superconductivity}

The type-I/type-II superconductivity of \NbGe~is shown as the orange zone in Fig.~\ref{fig: PD}. However, there are some difficulties in explaining the type-I/type-II superconductivity in \NbGe. The “two-component" theory suggests that in a multi-band superconductor, if there are two coherence lengths and the magnetic field penetration depth is between them, the type-I/type-II superconductivity is shown. Such theory has been successfully applied to explain the type-I/type-II superconductivity in MgB$_2$~\cite{Kortus2001} and ZrB$_{12}$~\cite{Lortz2005}. Nevertheless, there is no evidence that \NbGe~shows any multi-band superconducting feature based on current researches. Meanwhile, the IMS predicted in the single-band theory requires the condition that $\kappa$ is close to 1/$\sqrt{2}$. While, the $\kappa$ of \NbGe~estimated in Ref.~\cite{Lv2020}~is only 0.12, much smaller than 1/$\sqrt{2}$. 

Note that GL parameter $\kappa$ is defined as the ratio of two $T$-dependent characteristic lengths $\kappa = \lambda(T)/\xi(T)$~\cite{Tinkham2004}, and the penetration depth $\lambda$ also exhibits field dependence in some cases. For classic pure superconductors ($\kappa\ll1$) or dirty superconductors ($\kappa\gg1$), type-I or type-II superconductivity can be determined safely. But for type-I/type-II superconductors, we should examine the temperature and field dependence of $\kappa$ more carefully. We discuss the field and temperature dependence of $\kappa$ through further analysis of our TF-\musr~data in the mixed state of \NbGe~ in the following.

\subsubsection{Field dependence of $\lambda$}

\begin{figure}[ht]
	\begin{center}
		\includegraphics[clip=,width=8.5 cm]{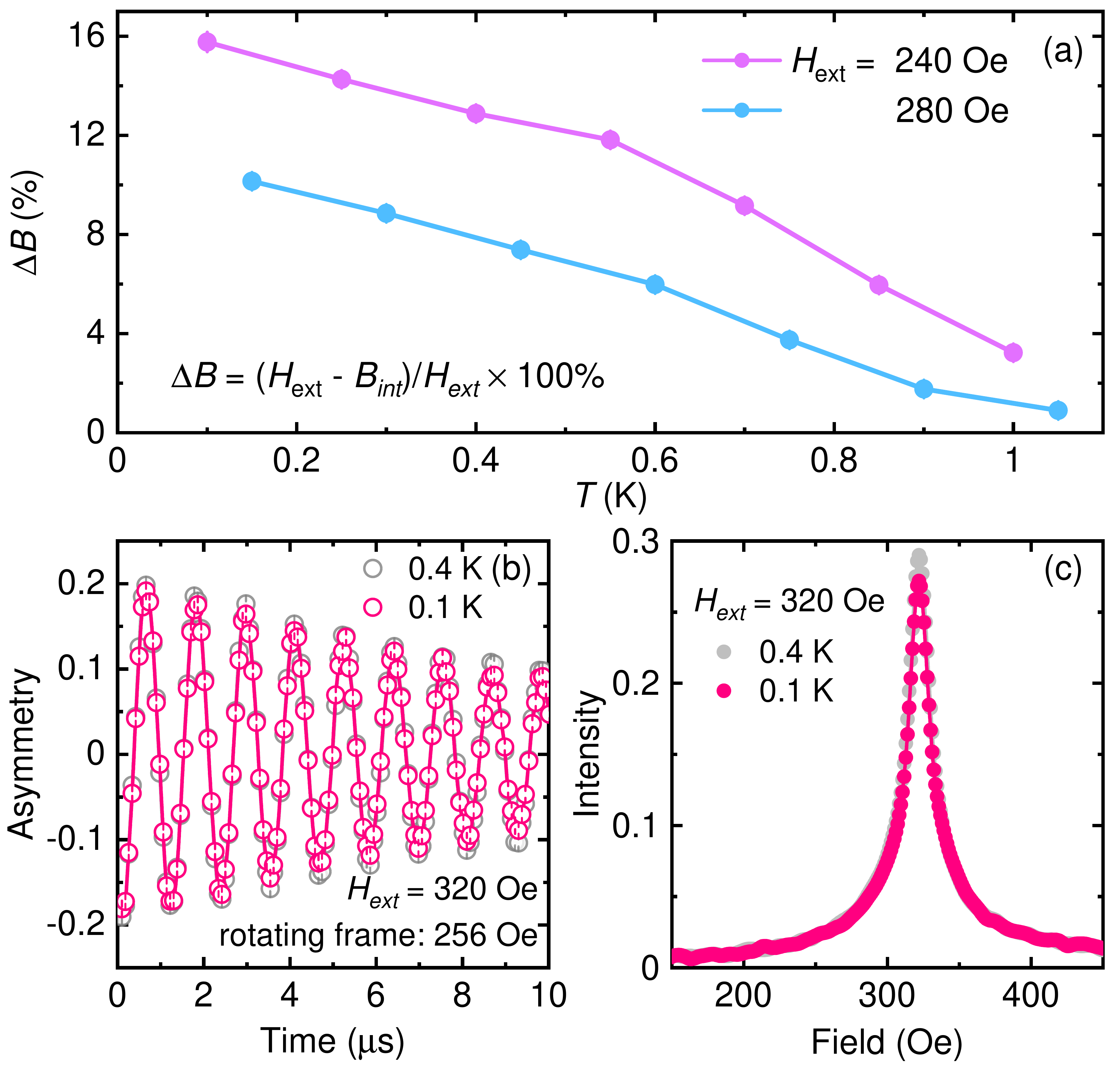}
		\caption{(a) $T$-dependence of $\Delta B$\% in the external field 240 Oe and 280 Oe. The difference between $H_{ext}$ and $B_{int}$ in the full $T$ range in $H_{ext}$ = 240 Oe is larger than that in $H_{ext}$ = 280 Oe. (b) TF-\musr~spectra from \NbGe~at $T$ = 0.1 K and 0.4 K with an external magnetic field of $H_{ext}$ = 320 Oe. Solid curves are fits to the data using Eq.~(\ref*{eq:TF2}). For clarity, the spectra are shown in a rotating reference frame corresponding to a field of 256 Oe. (c) FFTs curves of TF spectra in (b). }
		\label{fig: dis1}
	\end{center}
\end{figure}

When \NbGe~sample is in the vortex state, shown as the green regime in Fig.~\ref{fig: PD}, the function of TF-\musr~spectra is simplified as:
\begin{equation}
	\begin{split}
		\label{eq:TF2}
		Asy(t) = A_{bg}e^{-\lambda_{bg}t}\cos(\gamma_{\mu}H_{ ext}t+\phi)\\
		+ A_{vortex}\exp(-\frac{1}{2}\sigma_{ vortex}^{2}t^{2})\cos(\gamma_{\mu}B_{int}t+\phi).
	\end{split}
\end{equation}
As shown in Fig.~\ref{fig: TF2}, the field inside the sample $B_{int}$ formed by vortex lattice in the vortex state is smaller than the external field $H_{ext}$. For the same sample, the difference between $H_{ext}$ and $B_{int}$ can partially reflect the superconducting penetration depth. Generally, $H_{ext}$ and $B_{int}$ will become closer when there is a larger penetration depth $\lambda$. Fig.~\ref{fig: dis1} shows the bahavior of $\Delta B$\% = ($H_{ext}$ - $B_{int}$)/$H_{ext}$ $\times$ 100\% in different external fields. 

As shown in Fig.~\ref{fig: dis1}(a), the purple and blue curves represent $\Delta B$\% data in the external field $H_{ext}$ = 240 Oe and 280 Oe, respectively. All blue points appear below the purple points, which implies that the magnetic penetration depth $\lambda$ in $H_{ext}$ = 280 Oe is larger than that in $H_{ext}$ = 240~Oe. Meanwhile, as shown in Fig.~\ref{fig: dis1}(a) and (b), only slightly faster damping is observed at base temperature compared to 0.4 K with $H_{ext}$ = 320~Oe, and the magnetic fields in the mixed state are relatively close to $H_{ext}$. Therefore, the penetration depth $\lambda$ of \NbGe~increases under high fields, inducing the transition of superconductivity from type-I to type-II. At $T=0.3$~K, such crossover occurs at $H_{ext}=160$~Oe. This is consistent the previous MPCS experiment, in which the type-I and type-II superconductivity crossover at $T$ = 0.3~K originates from a magnetic field above 150~Oe~\cite{Zhang2021}.

\subsubsection{Temperature dependence of $\kappa$}

\begin{figure}[ht]
	\begin{center}
		\includegraphics[clip=,width=8.5 cm]{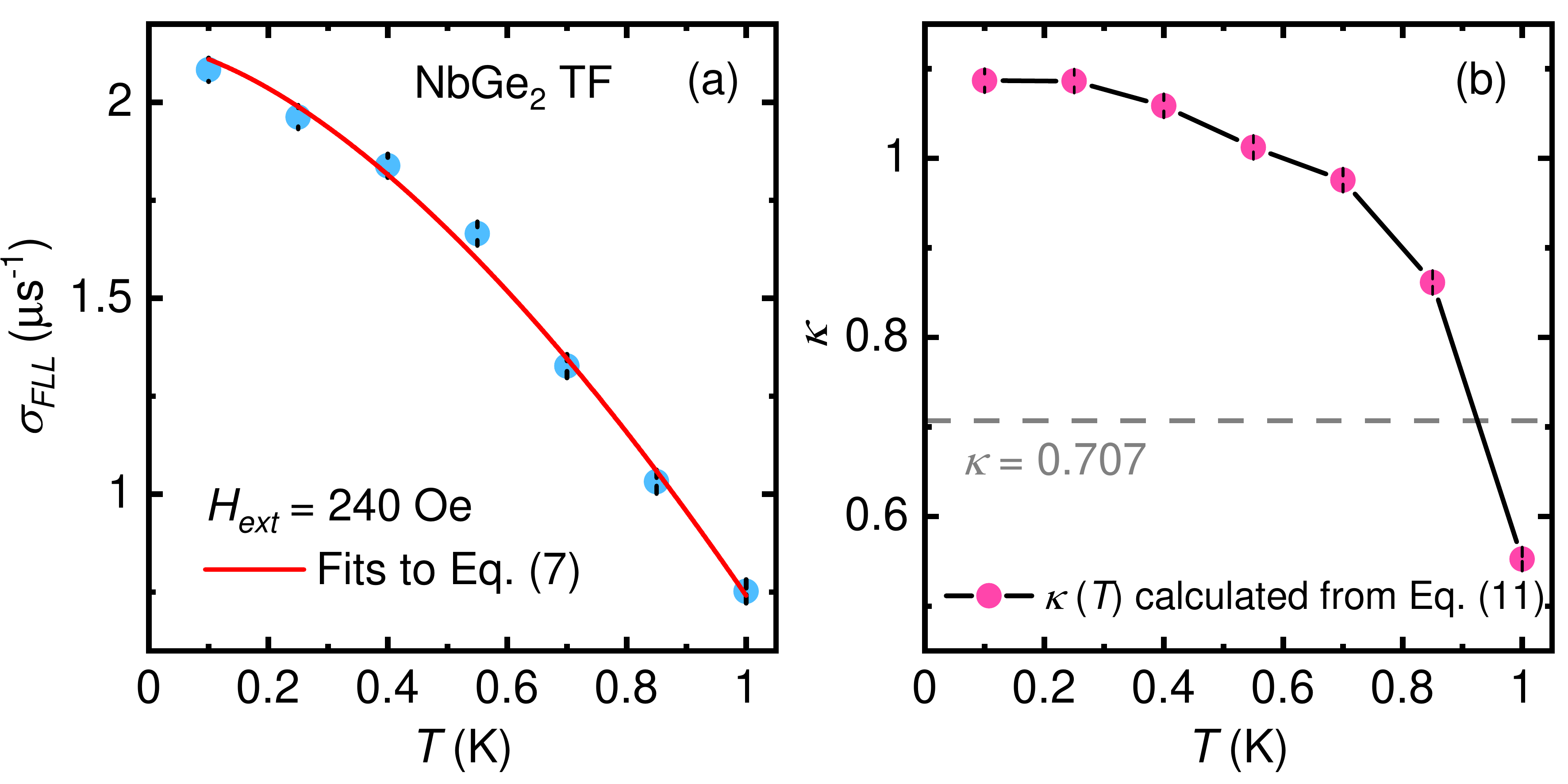}
		\caption{(a) Temperature dependence of  $\sigma_{FLL}$ from Eq.~(\ref{eq:FLL}) in \NbGe~for $H_{ext} = 240$~Oe. Solid curve is fit to Eq.~(\ref{eq:FLL-T}). (b) Temperature dependence of the GL parameter $\kappa$, which is calculated from the Eq.~(\ref{eq:kappa}). The grey dashed line marks $\kappa=1/\sqrt{2}$. }
		\label{fig: dis2}
	\end{center}
\end{figure}

In this section, we analyze TF data with $H_{ext} = 240$~Oe. From Eq.~(\ref{eq:TF2}), the temperature dependence of $\sigma_{mix}$ is obtained. The internal field distribution in the vortex state is the convolution of the flux-line lattice (FLL) and the nuclear dipolar moment of the sample:
\begin{equation}
	\label{eq:FLL}
	\sigma_{vortex}^2 = \sigma_{dip}^2 +  \sigma_{FLL}^2
\end{equation}
where $\sigma_{dip}$ is the contribution from the randomly oriented nuclear dipolar moments and $\sigma_{FLL}$ is the FLL contribution. In the normal state, $\sigma_{vortex}$ = $\sigma_{dip}$ is roughly field and temperature independent, and $\sigma_{dip}$ is fixed at the average value 0.265(5)~$\mathrm{\mu}$s$^{-1}$.  Fig.~\ref{fig: dis2}(a) shows the temperature dependence of $\sigma_{FLL}$.

The temperature dependence of the $\sigma_{FLL}$ can be described by a power-law function: 
\begin{equation}
	\label{eq:FLL-T}
	\sigma_{FLL}(T) = \sigma_{FLL}(0) [1-(T/T_\mathrm{c})^n]
\end{equation}
yielding \Tc~=~1.25(3) K and $n$~=~1.53(7). Noted that such fitting is not the phenomenological two-fluid model which has been widely used to analyze the $T$-evolution of the superfluid density~\cite{Broholm1990,Manzano2002,Takeshita2009} in type-II superconductors. The specific correlation between relaxation rates $\sigma_{FLL}$ and the superfluid density in \NbGe~needs to be determined by further studies.

Note that there is the following relationship between the relaxation rate $\sigma_{FLL}$ and the root mean square (rms) $B_{loc}^{rms}$ of the field distribution in the FLL:
\begin{equation}
	\label{eq:FLL-1}
	\sigma_{FLL} = \gamma_{\mu}B_{loc}^{rms}.
\end{equation}

For type-II superconductor samples with $(H_{ext}/H_{c2})^{1/2} \lesssim 1$, $B_{loc}^{rms}$ is directly related to GL parameter $\kappa$ by~\cite{Brandt2003}
\begin{equation}
	\label{eq:FLL-2}
	B_{loc}^{rms} = 0.172\frac{1 - b}{\kappa^2 - 0.069}H_{c2},
\end{equation}
where $b$ is the reduced applied field $b = H_{ext}/H_{c2}$, and the temperature dependence of $H_{c2}(T)$ of \NbGe~can be described as~\cite{Lv2020}:
\begin{equation}
	\label{eq:FLL-3}
	H_{c2}(T) = H_{c2}(0)\frac{1-(T/T_c)^2}{1+(T/T_c)^2}.
\end{equation}

In addition, from Ref.~\onlinecite{Brandt2003}, it can be estimated that the Eq.~(\ref{eq:FLL-2}) is valid when $b^{1/2} > 0.85$, which is $H_{ext}/H_{C2} > 0.72$. Our experimental conditions meet the applicable range of Eq.~(\ref{eq:FLL-2}).

From the above three equations, we can deduce the correlation between $\kappa$ and $\sigma_{FLL}$:
\begin{equation}
	\label{eq:kappa}
	\kappa(T) = [0.069 + 0.172\gamma_{\mu}\frac{H_{c2}(T) - H_{ext}}{\sigma_{FLL}(T)}]^{1/2}
\end{equation}

Using Eq.~(\ref{eq:kappa}), the temperature dependence of $\kappa$ with $H_{ext}$~=~240~Oe is shown in Fig.~\ref{fig: dis2}(b). The $\kappa$ of \NbGe~is negativlely correlated with temperature. When approaching \Tc, $\kappa$ decreases rapidly and the sample exhibits type-I superconductivity. When the temperature decreases, the $\kappa$ of \NbGe~sample falls within the range $0.5<\kappa<\sqrt{2}$, thus exhibiting type-I/type-II superconductivity below a characteristic temperature $T^*$. From the visual extension of the data in Fig.~\ref{fig: dis2}(b), it can be estimated that $\kappa$(0 K)~$\approx$~1.1, which is consistent with previous reports. In addition, it should be noted that when using Eq.~(\ref{eq:kappa}) to estimate $\kappa(T)$, if $T/$\Tc~or $H_{ext}/H_{c2}$ are very close to 1, then the values of $[H_{c2}(T) - H_{ext}]$ and $\sigma_{FLL}$ will be very small, resulting in a large uncertainty in $\kappa(T)$. Therefore, using TF-\musr~to estimate $\kappa$ is more accurate in low temperature and intermediate external field conditions.

The above discussion can help explain the type-I/type-II superconductivity in \NbGe~phenomenologically. However, the specific mechanism of this type-I/type-II superconductivity and what role ASOC caused by the NCS structure plays in it are still unresolved, and more theories and experiments are needed for further study.

\section{Conclusion}

In summary, we performed ZF and TF-\musr~experiments on polycrystalline \NbGe. The preservation of TRS is confirmed by ZF-\musr. TF-\musr~experiments were performed to map the phase diagram of \NbGe. The unconventional \musr~responses of the coexistence of type-I and type-II superconductivity, representing the type-I/type-II behavior, are shown in the phase diagram. Boundaries among type-I, type-II superconductivity, and their crossover are clearly delineated. From TF-\musr, it is concluded that the magnetic field penetration depth $\lambda$ and the external field $H_{ext}$ are positively correlated, while the GL parameter $\kappa$ and the temperature are negatively correlated, thus we phenomenologically explaining the appearance of type-I/type-II superconductivity in \NbGe.

\begin{acknowledgments}

We are grateful to the STFC and the ISIS Neutron and Muon Facility for beamtime (Grant No: RB2310278), the support of the ISIS Cryogenic group, and the staff of J-PARC DR (2020A0051) for their valuable help during the \musr~experiments. This research was funded by the National Key Research and Development Program of China (Grant Nos. 2022YFA1402203 and 2019YFA0308602), the National Science Foundation of China (Grant Nos. 12174065 and 12174334), and the Shanghai Municipal Science and Technology Major Project Grant, No.~2019SHZDZX01.
	
\end{acknowledgments}

%

\end{document}